# Unified Theory of Magnetoelastic Effects in B20 chiral magnets

Yangfan Hu[1,*], Biao Wang[1,**]

[1]Sino-French Institute of Nuclear Engineering and Technology, Sun Yat-Sen University, 510275 GZ, China

A magnetic skyrmion is a spin whirl with topological protection and high mobility to electric current. Intrinsic magnetoelastic coupling in chiral magnets permits manipulation of magnetic skyrmions and their lattice using mechanical loads, which is essential for developing future spintronics devices. It is found in experiments that the stability and deformation of skyrmions are sensitive to stresses, while the appearance of magnetic skyrmions in turn has a significant effect on the mechanical properties of the underlying material. However, a theory which explains these related phenomena within a unified framework is not seen. Here we construct a thermodynamic model for B20 helimagnets incorporating a magnetoelastic functional with necessary higher order interactions derived by group theory. Within the model, we establish the methodology to calculate the phase diagram and equilibrium properties of helimagnets under coupled temperature-magneto-elastic field. Applying the model to bulk MnSi, we calculate the temperature-magnetic field phase diagram under stress-free condition and its variation when uniaxial compression is applied. We also calculate the variation of all the elastic constants with magnetic field. The results obtained agree quantitatively with corresponding experiments. Our model provides a reliable basis for further theoretical studies concerning any magnetoelastic related phenomena in chiral magnets.

## I. Introduction

Last few years have witnessed a revival of interest in chiral magnets such as MnSi, $Fe_{0.5}Co_{0.5}Si$, and FeGe, due to experimental observation of a new chiral modulated magnetic state, commonly referred to as the skyrmion lattice phase[1-3]. The skyrmion lattice phase can be understood as crystallization of isolated skyrmions, the latter of which are stabilized by the antisymmetric Dzyaloshinskii-Moriya (DM) interaction[4-6]. A skyrmion is attractive for its emergent electromagnetic properties such as spin motive force[7-9] and topological Hall effect[10, 11]. The stability and appearance of skyrmions are sensitive to material size, shape[12-16] and various kinds of external fields[17-19]. Since the critical current density required to drive the motion of skyrmions is much lower than that for a magnetic domain wall[20, 21], magnetic skyrmions are promising candidate for realization of the next generation spintronic devices.

Magnetoelastic coupling in chiral magnets permits interaction between skyrmions and the elastic fields of the underlying material, which leads to occurrence of profound magnetoelastic phenomena. Firstly, application of mechanical loads can affect or even stabilize the skyrmion lattice. Through a theoretical model developed upon the Landau-Ginzburg functional, A. B. Butenko et al.[22] find that uniaxial distortion stabilizes the skyrmion lattice in a broad range of thermodynamical parameters in cubic noncentrosymmetric ferromagnets, and further argue that this mechanism is responsible for the formation of skyrmion states observed in thin layers of $Fe_{0.5}Co_{0.5}Si$[2].

*Corresponding author. E-mail: huyf3@mail.sysu.edu.cn (Yangfan Hu),
**Corresponding author. E-mail: wangbiao@mail.sysu.edu.cn (Biao Wang).

Stabilization of skyrmion lattice is also observed in epitaxially grown FeGe thin films[23] and MnSi thin films[24], where the magnetoelastic effects should be significant due to the misfit strains exerted by the substrate. Later, it is observed in bulk MnSi that uniaxial tension affects the stability area of the skyrmion lattice phase in the phase diagram[25]. Secondly, the existence of magnetic skyrmions in turn affects elastic behaviour of the material. Through ultrasonic studies[26, 27], it is found that appearance of skyrmion lattice in chiral magnets is accompanied by a jump of elastic stiffness. Finally, deformation of the skyrmions and their lattice is strongly coupled with deformation of the underlying materials. It is proved theoretically that presence of skyrmions lead to nontrivial localized elastic fields[28, 29] when the materials are free from external forces. And when uniaxial tension is applied, the skyrmion lattice in FeGe thin film[30] undergoes dramatic distortion two orders of magnitude larger than that of the underlying material.

To clarify the physical mechanism behind, it is significant to establish a theory that can treat these different aspects of magnetoelastic phenomena within a unified framework. Previous theoretical studies on magnetoelastic coupling in MnSi or other chiral magnets mainly fall into two categories: one part is developed upon the magnetostriction theory constructed for ferromagnets with cubic symmetry (hereafter referred to as K theory)[31], while the other part is developed upon a Landau type mean-field model constructed for the specific spin-density-wave phase of MnSi (hereafter referred to as P theory)[32]. After thorough investigation, we find both theories to be oversimplified to explain skyrmion-related magnetoelastic phenomena: the variation of the elastic constants with the magnetic field examined for MnSi[26, 27] in the skyrmion lattice phase cannot be understood within both theory. Recently, we became aware of a paper by Zhang and Nagaosa[33] addressing the ultrasonic elastic responses in a monopole lattice using a extended spin-wave theory concerning magnon-phonon interaction. While such a microscopic model provides a deeper understanding of the origin of magnetoelastic coupling in chiral magnets, it is found that their model is more applicable to MnGe than MnSi. A possible reason is that the magnetoelastic Hamiltonian used in the model is constructed in a most simple form instead of a comprehensive description derived upon symmetry consideration. Besides, the effect of transverse acoustic waves is not concerned in the model, so that it cannot be used to analyze the variation of shear elastic constants $C_{44}$ and $C_{66}$ with the magnetic field.

In this paper, we formulate a thermodynamic model for B20 helimagnets incorporating a comprehensive magnetoelastic functional. The magnetoelastic functional is derived for the first time based on symmetry consideration of B20 helimagnets, where all necessary higher order interactions are incorporated, so that the variation of all elastic constants of MnSi with external magnetic field observed in experiments can be quantitatively explained. The model is fundamental for studying any skyrmion-related magnetoelastic phenomena. Here we explain two basic utilities of the model: a) calculation of temperature-magnetic field phase diagram of any B20 compounds at any applied mechanical loads; b) calculation of equilibrium properties for helimagnets at any given temperature, magnetic field, and mechanical loads. We apply the theory to bulk MnSi to evaluate the effect of magnetoelastic coupling on the temperature-magnetic field phase diagram, the equilibrium magnetization, and the elastic constants. By doing so, a complete set of thermodynamic parameters is determined for MnSi.

## II. Formulation of the magnetoelastic free energy density functional

The space group of B20 compounds is denoted by P2$_1$3, which corresponds to point group T or the tetrahedral group. According to the symmetry requirement of point group T, we propose the following magnetoelastic energy density functional for B20 compounds, which is written in a rescaled form as:

$$\widetilde{w}_{me}(\varepsilon_{ij}, M_k, M_{l,m}) = \widetilde{w}_{me0} + \widetilde{w}_{me1} + \widetilde{w}_{me2}, \qquad (1)$$

where

$$\widetilde{w}_{me0} = \widetilde{K}m^2\varepsilon_{ii} + \widetilde{L}_1(m_1^2\varepsilon_{11} + m_2^2\varepsilon_{22} + m_3^2\varepsilon_{33}) + \widetilde{L}_2(m_3^2\varepsilon_{11} + m_1^2\varepsilon_{22} + m_2^2\varepsilon_{33}) \\ + \widetilde{L}_3(m_1 m_2 \gamma_{12} + m_1 m_3 \gamma_{13} + m_2 m_3 \gamma_{23}), \qquad (2)$$

$$\widetilde{w}_{me1} = \sum_{i=1}^{6} \widetilde{L}_{0i}\tilde{f}_{0i}, \qquad (3)$$

$$\widetilde{w}_{me2} = \sum_{i=1}^{3} \widetilde{L}_{2i}\tilde{f}_{2i}, \qquad (4)$$

and

$$\begin{aligned}
\tilde{f}_{O1} &= \varepsilon_{11}(m_{1,2}m_3 - m_{1,3}m_2) + \varepsilon_{22}(m_{2,3}m_1 - m_{2,1}m_3) + \varepsilon_{33}(m_{3,1}m_2 - m_{3,2}m_1), \\
\tilde{f}_{O2} &= \varepsilon_{11}(m_{3,1}m_2 - m_{2,1}m_3) + \varepsilon_{22}(m_{1,2}m_3 - m_{3,2}m_1) + \varepsilon_{33}(m_{2,3}m_1 - m_{1,3}m_2), \\
\tilde{f}_{O3} &= \varepsilon_{11}m_1(m_{2,3} - m_{3,2}) + \varepsilon_{22}m_2(m_{3,1} - m_{1,3}) + \varepsilon_{33}m_3(m_{1,2} - m_{2,1}), \\
\tilde{f}_{O4} &= \gamma_{23}(m_{1,3}m_3 - m_{1,2}m_2) + \gamma_{13}(m_{2,1}m_1 - m_{2,3}m_3) + \gamma_{12}(m_{3,2}m_2 - m_{3,1}m_1), \\
\tilde{f}_{O5} &= \gamma_{23}(m_{3,1}m_3 - m_{2,1}m_2) + \gamma_{13}(m_{1,2}m_1 - m_{3,2}m_3) + \gamma_{12}(m_{2,3}m_2 - m_{1,3}m_1), \\
\tilde{f}_{O6} &= \gamma_{23}m_1(m_{3,3} - m_{2,2}) + \gamma_{13}m_2(m_{1,1} - m_{3,3}) + \gamma_{12}m_3(m_{2,2} - m_{1,1}),
\end{aligned} \qquad (5)$$

$$\begin{aligned}
\tilde{f}_{21} &= \gamma_{23}\gamma_{12}m_1 m_3 + \gamma_{23}\gamma_{13}m_1 m_2 + \gamma_{12}\gamma_{13}m_2 m_3, \\
\tilde{f}_{22} &= \gamma_{23}^2 m_1^2 + \gamma_{12}^2 m_3^2 + \gamma_{13}^2 m_2^2, \\
\tilde{f}_{23} &= m^2(\gamma_{23}^2 + \gamma_{12}^2 + \gamma_{13}^2).
\end{aligned} \qquad (6)$$

In eq. (1), the rescaled magnetoelastic energy density $\widetilde{w}_{me}$ is expressed as a functional of the normal elastic strains $\varepsilon_{11}, \varepsilon_{22}, \varepsilon_{33}$, the engineering shear strains $\gamma_{23}, \gamma_{13}, \gamma_{12}$, the rescaled magnetization $\mathbf{m} = [m_1, m_2, m_3]^T$ and its first order partial derivatives $m_{i,j}, (i, j = 1, 2, 3)$. In eqs. (2-4), $m^2 = m_1^2 + m_2^2 + m_3^2$, and $\widetilde{K}, \widetilde{L}_1, \widetilde{L}_2, \widetilde{L}_3, \widetilde{L}_{0i}, (i = 1, 2, \ldots, 6)$ and $\widetilde{L}_{2i}, (i = 1, 2, 3)$ are thermodynamic parameters characterizing different orders of magnetoelastic interactions. One should notice that the engineering shear strains are related to the shear strains by $\gamma_{12} = 2\varepsilon_{12}, \gamma_{13} = 2\varepsilon_{13}$, and $\gamma_{23} = 2\varepsilon_{23}$.

Putting $\widetilde{L}_2 = \widetilde{K} = 0$ in eq. (2), $\widetilde{w}_{me0}$ reduces to the magnetoelastic interactions defined in the K theory. Changing the first term on the right hand side of eq. (2) as $\widetilde{L}_1[m^2\varepsilon_{ii} - (m_3^2\varepsilon_{11} + m_1^2\varepsilon_{22} + m_2^2\varepsilon_{33}) - (m_2^2\varepsilon_{11} + m_3^2\varepsilon_{22} + m_1^2\varepsilon_{33})]$ and merging similar terms, $\widetilde{w}_{me0}$ reduces to the functional used by the P theory. The discrepancy between the K theory and the P theory in describing magnetoelastic interactions derives from the symmetry requirement of different point groups in cubic systems, which was well summarized long ago[34].

Neither the K theory nor the P theory is sufficient to explain the complex variation of elastic coefficients with external magnetic field discovered in experiments of B20 compounds[26, 27], for which higher order interactions $\widetilde{w}_{me1}$ and $\widetilde{w}_{me2}$ are introduced in eq. (1). $\widetilde{w}_{me1}$ is needed when explaining the discrepancy between the elastic constants $C_{11}$ and $C_{33}$ in the skyrmion phase observed in ultrasonic experiment of MnSi. On the other hand, $w_{me2}$ is needed when explaining the variation of $C_{44}$ and $C_{66}$ with the magnetic field[27]. One should notice that the two parts of functional have already been simplified, where the details are described in Appendix A.

## III. Extended micromagnetic model incorporating magnetoelastic interaction

The Helmholtz free energy density for cubic helimagnets suffering coupled temperature-magneto-elastic field can be derived by incorporating the magnetoelastic functional developed in section II in the Ginzburg-Landau functional for chiral magnets with cubic symmetry[1, 4, 5, 35, 36]. We present the Helmholtz free energy density of the system in a rescaled form as

$$\widetilde{w}(\mathbf{m}, \varepsilon_{ij}) = \sum_{i=1}^{3} \left(\frac{\partial \mathbf{m}}{\partial r_i}\right)^2 + 2\mathbf{m} \cdot (\nabla \times \mathbf{m}) - 2\mathbf{b} \cdot \mathbf{m} + t\mathbf{m}^2 + \mathbf{m}^4 + \widetilde{w}_{an} + \widetilde{w}_{el} + \widetilde{w}_{me}, \quad (7)$$

where

$$\widetilde{w}_{an} = \sum_{i=1}^{3} \left[\tilde{A}_e \left(\frac{\partial m_i}{\partial r_i}\right)^2 + \tilde{A}_c m_i^4\right], \quad (8)$$

$$\widetilde{w}_{el} = \frac{1}{2}\tilde{C}_{11}(\varepsilon_{11}^2 + \varepsilon_{22}^2 + \varepsilon_{33}^2) + \tilde{C}_{12}(\varepsilon_{11}\varepsilon_{22} + \varepsilon_{11}\varepsilon_{33} + \varepsilon_{22}\varepsilon_{33})$$
$$+ \frac{1}{2}\tilde{C}_{44}(\gamma_{12}^2 + \gamma_{13}^2 + \gamma_{23}^2) \quad (9)$$

denote respectively the rescaled anisotropic energy density and the rescaled elastic energy density, and $\widetilde{w}_{me}$ is obtained by rescaling the magnetoelastic free energy density developed in section II. In eq. (7), $\widetilde{w}$ is given as a functional of the rescaled magnetization vector $\mathbf{m}$ and the elastic strains $\varepsilon_{ij}$ at given rescaled temperature $t$ and rescaled magnetic field $\mathbf{b}$. Such a rescaled form reduces the number of effective thermodynamic parameters, and provides a material-independent theoretical framework to discuss the effect of magnetoelastic coupling. The rescaling process and the definition of all quantities with a wavy overline are given in Appendix B. Eq. (7) is fundamental to study various kinds of phenomena that occur in chiral magnets suffering coupled temperature-magneto-elastic field. Here we discuss two kinds of basic utilities as follow.

### A. Temperature-magnetic field phase diagram calculation for helimagnets suffering mechanical loads

To proceed, the first step is to solve the elastic strains at given temperature, magnetic field, and mechanical loads, which differs for different kinds of mechanical boundary conditions. For displacement boundary condition where the displacements are fixed at the boundaries as $u_{i0}$, the elastic strains are fully determined by the boundary condition as $\varepsilon_{ij} = \varepsilon_{ij}(u_{i0})$. For stress boundary condition, the stresses $\sigma_{ij}$ are determined by solving an elasticity problem using stress methods (e.g., the method of Airy stress function). Then the elastic strains $\varepsilon_{ij}$ can be solved from the constitutive equations $\tilde{\sigma}_{ij} = \frac{\partial \widetilde{w}(\mathbf{m}, \varepsilon_{ij})}{\partial \varepsilon_{ij}}$ as

$$\begin{aligned}
\tilde{\sigma}_{11} &= \tilde{C}_{11}(\varepsilon_{11} - \varepsilon_{11}^*) + \tilde{C}_{12}(\varepsilon_{22} - \varepsilon_{22}^* + \varepsilon_{33} - \varepsilon_{33}^*), \\
\tilde{\sigma}_{22} &= \tilde{C}_{11}(\varepsilon_{22} - \varepsilon_{22}^*) + \tilde{C}_{12}(\varepsilon_{11} - \varepsilon_{11}^* + \varepsilon_{33} - \varepsilon_{33}^*), \\
\tilde{\sigma}_{33} &= \tilde{C}_{11}(\varepsilon_{33} - \varepsilon_{33}^*) + \tilde{C}_{12}(\varepsilon_{11} - \varepsilon_{11}^* + \varepsilon_{22} - \varepsilon_{22}^*), \\
\begin{bmatrix} \tilde{\sigma}_{23} \\ \tilde{\sigma}_{13} \\ \tilde{\sigma}_{12} \end{bmatrix} &= \mathbf{D} \begin{bmatrix} \gamma_{23} - \gamma_{23}^* \\ \gamma_{13} - \gamma_{13}^* \\ \gamma_{12} - \gamma_{12}^* \end{bmatrix},
\end{aligned} \quad (10)$$

where $\tilde{\sigma}_{ij}$ denotes the rescaled stress components, and $\varepsilon_{ij}^*$ and $\gamma_{ij}^*$, the eigenstrains, are related to the rescaled magnetization by

$$\varepsilon_{11}^* = K^*m^2 - L_1^*m_1^2 - L_2^*m_3^2 + L_{O1}^*(m_3m_{1,2} - m_2m_{1,3}) + L_{O2}^*(m_3m_{2,1} - m_2m_{3,1}) + L_{O3}^*m_1(m_{2,3} - m_{3,2}),$$
$$\varepsilon_{22}^* = K^*m^2 - L_1^*m_2^2 - L_2^*m_1^2 + L_{O1}^*(m_1m_{2,3} - m_3m_{2,1}) + L_{O2}^*(m_1m_{3,2} - m_3m_{1,2}) + L_{O3}^*m_2(m_{3,1} - m_{1,3}),$$
$$\varepsilon_{33}^* = K^*m^2 - L_1^*m_3^2 - L_2^*m_2^2 + L_{O1}^*(m_2m_{3,1} - m_1m_{3,2}) + L_{O2}^*(m_2m_{1,3} - m_1m_{2,3}) + L_{O3}^*M_3(m_{1,2} - m_{2,1}), \quad (11)$$
$$\begin{bmatrix} \gamma_{23}^* \\ \gamma_{13}^* \\ \gamma_{12}^* \end{bmatrix} = \mathbf{D}^{-1} \begin{bmatrix} \sigma_{23}^* \\ \sigma_{13}^* \\ \sigma_{12}^* \end{bmatrix},$$

and

$$\mathbf{D} = \begin{bmatrix} (\tilde{C}_{44}^* + 2\tilde{L}_{22}m_1^2) & \tilde{L}_{21}m_1m_2 & \tilde{L}_{21}m_1m_3 \\ \tilde{L}_{21}m_1m_2 & (\tilde{C}_{44}^* + 2\tilde{L}_{22}m_2^2) & \tilde{L}_{21}m_2m_3 \\ \tilde{L}_{21}m_1m_3 & \tilde{L}_{21}m_2m_3 & (\tilde{C}_{44}^* + 2\tilde{L}_{22}m_3^2) \end{bmatrix}, \quad (12)$$

$$\sigma_{23}^* = -\tilde{L}_3 m_2 m_3 + \tilde{L}_{O6} m_1(m_{2,2} - m_{3,3}) + m_2(\tilde{L}_{O4}m_{1,2} + \tilde{L}_{O5}m_{2,1}) - m_3(\tilde{L}_{O4}m_{1,3} + \tilde{L}_{O5}m_{3,1}),$$
$$\sigma_{13}^* = -\tilde{L}_3 m_1 m_3 + \tilde{L}_{O6} m_2(m_{3,3} - m_{1,1}) + m_3(\tilde{L}_{O4}m_{2,3} + \tilde{L}_{O5}m_{3,2}) - m_1(\tilde{L}_{O4}m_{2,1} + \tilde{L}_{O5}m_{1,2}),$$
$$\sigma_{12}^* = -\tilde{L}_3 m_1 m_2 + \tilde{L}_{O6} m_3(m_{1,1} - m_{2,2}) + m_1(\tilde{L}_{O4}m_{3,1} + \tilde{L}_{O5}m_{1,3}) - m_2(\tilde{L}_{O4}m_{3,2} + \tilde{L}_{O5}m_{2,3}).$$

In eqs. (11, 12), the parameters with a superscript "*" are defined as $K^* = \frac{-\tilde{C}_{11}\tilde{K} + \tilde{C}_{12}(\tilde{K} + \tilde{L}_1 + \tilde{L}_2)}{(\tilde{C}_{11} - \tilde{C}_{12})(\tilde{C}_{11} + 2\tilde{C}_{12})}$, $L_1^* = \frac{\tilde{L}_1}{(\tilde{C}_{11} - \tilde{C}_{12})}$, $L_2^* = \frac{\tilde{L}_2}{(\tilde{C}_{11} - \tilde{C}_{12})}$, $L_{O1}^* = \frac{-\tilde{C}_{11}\tilde{L}_{O1} + \tilde{C}_{12}(-\tilde{L}_{O1} + \tilde{L}_{O2} + \tilde{L}_{O3})}{(\tilde{C}_{11} - \tilde{C}_{12})(\tilde{C}_{11} + 2\tilde{C}_{12})}$, $L_{O2}^* = \frac{\tilde{C}_{11}\tilde{L}_{O2} - \tilde{C}_{12}(\tilde{L}_{O1} - \tilde{L}_{O2} + \tilde{L}_{O3})}{(\tilde{C}_{11} - \tilde{C}_{12})(\tilde{C}_{11} + 2\tilde{C}_{12})}$, $L_{O3}^* = \frac{\tilde{C}_{12}(\tilde{L}_{O1} + \tilde{L}_{O2}) - (\tilde{C}_{11} + \tilde{C}_{12})\tilde{L}_{O3}}{(\tilde{C}_{11} - \tilde{C}_{12})(\tilde{C}_{11} + 2\tilde{C}_{12})}$, $\tilde{C}_{44}^* = \tilde{C}_{44} + 2\tilde{L}_{23}m^2$. For chiral magnetic states, the solution of elastic strains contains a homogeneous part and a periodic part: $\varepsilon_{ij} = \bar{\varepsilon}_{ij}(\mathbf{m}, \tilde{\sigma}_{ij}) + \hat{\varepsilon}_{ij}(\mathbf{m}, \tilde{\sigma}_{ij})$, where $\bar{\varepsilon}_{ij}(\mathbf{m}, \tilde{\sigma}_{ij}) = \frac{1}{V}\int_V \varepsilon_{ij}dV$. Here $\bar{\varepsilon}_{ij}(\mathbf{m}, \tilde{\sigma}_{ij})$ can be solved by taking volume average of eq. (10), while $\hat{\varepsilon}_{ij}(\mathbf{m}, \tilde{\sigma}_{ij})$ can be derived by solving an eigenstrain problem[29]. For mixed boundary condition, we generally have after deduction $\varepsilon_{ij} = \varepsilon_{ij}(u_{i0}, \mathbf{m}, \tilde{\sigma}_{ij})$, where $u_{i0}$ is the displacement prescribed at part of the boundary and $\tilde{\sigma}_{ij}$ are the stresses solved using the stress boundary condition prescribed at the other part of the boundary. In all three cases, the elastic strains can generally be written as functions of the rescaled magnetization $\mathbf{m}$: $\varepsilon_{ij} = \varepsilon_{ij}(\mathbf{m})$.

In the second step, we substitute $\varepsilon_{ij} = \varepsilon_{ij}(\mathbf{m})$ derived above into eq. (7), which expresses the rescaled free energy density as a mere functional of $\mathbf{m}$. Then we consider a specific magnetic phase, which describes $\mathbf{m}$ by a certain mathematical expression. For chiral magnets, the known magnetic phases include: i) the general conical phase

$$\mathbf{m}_{gconical} = \begin{bmatrix} \frac{1}{2}(\cos\theta + 1) & \frac{1}{2}(\cos\theta - 1) & \frac{\sqrt{2}}{2}\sin\theta \\ \frac{1}{2}(\cos\theta - 1) & \frac{1}{2}(\cos\theta + 1) & \frac{\sqrt{2}}{2}\sin\theta \\ -\frac{\sqrt{2}}{2}\sin\theta & -\frac{\sqrt{2}}{2}\sin\theta & \cos\theta \end{bmatrix} \begin{bmatrix} m_q\cos(\mathbf{q}\cdot\mathbf{r}) \\ m_q\sin(\mathbf{q}\cdot\mathbf{r}) \\ m_3 \end{bmatrix}, \quad (13)$$

where $\theta$ denotes the angle between the direction of $\mathbf{q}$ and $[0\ 0\ 1]^T$, $\mathbf{q} = q\left[\frac{\sqrt{2}}{2}\sin\theta \quad \frac{\sqrt{2}}{2}\sin\theta \quad \cos\theta\right]^T$. Here it is pre-assumed that $\mathbf{q}$ always lies in the $(1\bar{1}0)$ plane, which is determined by the easy axis of the material $[1\ 1\ 1]^T$ and the direction of magnetic field $[0\ 0\ 1]^T$. It reduces to ii) the conical phase when $\theta = 0$, which gives

$$\mathbf{m}_{conical} = [m_q\cos(qr_3) \quad m_q\sin(qr_3) \quad m_3]^T. \quad (14)$$

Eq. (13) reduces to iii) the general helical phase $\mathbf{m}_{ghelical}$ when $m_3 = 0$. Eq. (14) reduces to iv)

the helical phase $\mathbf{m}_{helical}$ when $m_3 = 0$, and it reduces to v) the ferromagnetic phase $\mathbf{m}_{ferro}$ when $m_q = 0$. vi) The skyrmion lattice phase within the $n^{th}$ order Fourier representation[37]:

$$\mathbf{m}_{Fn} = \mathbf{m}_0 + \sum_{i=1}^{n}\sum_{j=1}^{n_i} \mathbf{m}_{\mathbf{q}_{ij}} e^{i\mathbf{q}_{ij}\cdot\mathbf{r}}, \qquad (15)$$

where $|\mathbf{q}_{i1}| = |\mathbf{q}_{i2}| = |\mathbf{q}_{i3}| = \cdots = s_i q$, $|\mathbf{m}_{\mathbf{q}_{i1}}| = |\mathbf{m}_{\mathbf{q}_{i2}}| = |\mathbf{m}_{\mathbf{q}_{i3}}| = \cdots = m_{qi}$, $|\mathbf{q}_{1j}| < |\mathbf{q}_{2j}| < |\mathbf{q}_{3j}| < \cdots$, and $n_i$ denotes the number of reciprocal vectors whose modulus equals to $s_i q$. Here $s_i$ is a positive sequence of number that increases with $i$, and $\mathbf{m}_{\mathbf{q}_{ij}}$ can be expanded as

$$\mathbf{m}_{\mathbf{q}_{ij}} = c_{i1}\mathbf{P}_{ij1} + c_{i2}\mathbf{P}_{ij2} + c_{i3}\mathbf{P}_{ij3}, \qquad (16)$$

where

$$\mathbf{P}_{ij1} = \frac{1}{\sqrt{2}s_i q}[-iq_{ijy}, iq_{ijx}, s_i q]^T, \mathbf{P}_{ij2} = \frac{1}{s_i q}[q_{ijx}, q_{ijy}, 0]^T, \mathbf{P}_{ij3} = \frac{1}{\sqrt{2}s_i q}[iq_{ijy}, -iq_{ijx}, s_i q]^T \quad (17)$$

and $\mathbf{q}_{ij} = [q_{ijx} \quad q_{ijy} \quad 0]^T$. Eq. (15) is constructed upon the hexagonal symmetry of the skyrmion crystal. When distortion of the skyrmion crystal is considered, the emergent elastic strains have to be introduced in the Fourier representation[38]. By setting $n = 1$ and $c_{12} = c_{13} = 0$, eq. (15) reduces to the triple-Q representation which can be written without loss of generality as

$$\mathbf{m}_{tripleQ} = \begin{bmatrix} 0 \\ 0 \\ m_0 \end{bmatrix} + \sqrt{2}m_{q1}\left\{ \begin{bmatrix} \sin(\mathbf{q}_{11}\mathbf{r}) \\ 0 \\ \cos(\mathbf{q}_{11}\mathbf{r}) \end{bmatrix} + \begin{bmatrix} -\frac{1}{2}\sin(\mathbf{q}_{12}\mathbf{r}) \\ \frac{\sqrt{3}}{2}\sin(\mathbf{q}_{12}\mathbf{r}) \\ \cos(\mathbf{q}_{12}\mathbf{r}) \end{bmatrix} + \begin{bmatrix} -\frac{1}{2}\sin(\mathbf{q}_{13}\mathbf{r}) \\ -\frac{\sqrt{3}}{2}\sin(\mathbf{q}_{13}\mathbf{r}) \\ \cos(\mathbf{q}_{13}\mathbf{r}) \end{bmatrix} \right\}. \quad (18)$$

After specifying a magnetic phase, one solves the magnetization that minimizes the averaged free energy density $\overline{w}(\mathbf{m}, \varepsilon_{ij}) = \frac{1}{V}\int \widetilde{w}(\mathbf{m}, \varepsilon_{ij}) dV$. For example, if we consider the skyrmion lattice phase within the triple-Q representation, we minimize $\overline{w}(\mathbf{m}_{tripleQ}) = \overline{w}(m_0, m_{q1}, q)$ which determines the independent variables $m_0, m_{q1}, q$ and the minimized averaged free energy density $\overline{w}_{tripleQ}$.

In the third step, we repeat the free energy minimization process mentioned above for all possible magnetic phases. The equilibrium magnetic state is the one that yields the smallest averaged free energy density.

In the last step, we change the temperature and magnetic field, and then repeat the process mentioned above to determine the equilibrium magnetic state at the new condition. The temperature-magnetic field phase diagram is obtained after all points in the phase diagram are considered. A phase diagram of any two parameters can be derived in the same way if we change the temperature and magnetic field to two new parameters of interest.

**B. Equilibrium properties for helimagnets concerning magnetoelastic coupling**

To proceed, the first step is to determine the equilibrium magnetic state at given temperature, magnetic field, and mechanical loads, using the method introduced above in part A. By doing so, we obtain the value of all independent variables at the given condition.

The second step is to decide which kind of equilibrium properties are to be discussed, and clarify their thermodynamic definition (e.g., the rescaled elastic stiffness $\tilde{C}_{ijkl} = \frac{\partial \tilde{\sigma}_{ij}}{\partial \varepsilon_{kl}} = \frac{\partial^2 \overline{w}}{\partial \varepsilon_{ij}\partial \varepsilon_{kl}}$). Then we determine the work conjugates of all independent variables of the equilibrium magnetic state. The equilibrium property of interest is to be calculated at given temperature and corresponding work conjugates of the independent variables.

The third step is to solve the analytical expression of the equilibrium properties from the averaged free energy density by using the method of Jacobian transformation in thermodynamics[39]. The analytical expressions are generally lengthy and symbolic computation programs are needed for the deduction. When the analytical expressions are derived, we substitute the values of all independent variables obtained in the first step to calculate the values of the equilibrium properties.

Finally, we change the magnetic field and repeat the process above. The variation of the equilibrium properties of interest with the magnetic field is then obtained. Here the magnetic field can be replaced by another parameter, and the variation of the equilibrium properties with the parameter can be obtained in the same way.

As an example, we derive for the first time the elastic stiffness in the skyrmion lattice phase at given temperature, magnetic field and mechanical loads. At rescaled temperature $t$, rescaled magnetic field $\mathbf{b} = [0 \quad 0 \quad b]^T$, and elastic constrains $\varepsilon_{ij} = 0$, the equilibrium magnetic state is found to be the skyrmion lattice phase, described within the n$^{\text{th}}$ order Fourier representation $\mathbf{m} = \mathbf{m}_{Fn}(m_0, c_{11}, c_{12}, c_{13}, c_{21}, c_{22}, c_{23}, \ldots, c_{n1}, c_{n2}, c_{n3})$. The work conjugate of $m_0$ is found to be $b$, while the work conjugates of $c_{11}, c_{12}, c_{13}, c_{21}, c_{22}, c_{23}, \ldots, c_{n1}, c_{n2}$, and $c_{n3}$ are denoted by $b_{11}, b_{12}, b_{13}, b_{21}, b_{22}, b_{23}, \ldots, b_{n1}, b_{n2}$, and $b_{n3}$. The elastic constants at given condition can be derived from

$$\left(\tilde{C}_{ijkl}\right)_{t,b,b_{i1},b_{i2},b_{i3},\varepsilon_{ij}=0} = \left[\frac{\left[\frac{\partial(\tilde{\sigma}_{ij}, b, b_{11}, b_{12}, b_{13}, b_{21}, b_{22}, b_{23}, \ldots, b_{n1}, b_{n2}, b_{n3})}{\partial(\varepsilon_{kl}, m_0, c_{11}, c_{12}, c_{13}, c_{21}, c_{22}, c_{23}, \ldots, c_{n1}, c_{n2}, c_{n3})}\right]}{\left[\frac{\partial(\varepsilon_{kl}, b, b_{11}, b_{12}, b_{13}, b_{21}, b_{22}, b_{23}, \ldots, b_{n1}, b_{n2}, b_{n3})}{\partial(\varepsilon_{kl}, m_0, c_{11}, c_{12}, c_{13}, c_{21}, c_{22}, c_{23}, \ldots, c_{n1}, c_{n2}, c_{n3})}\right]}\right]_{t,b,\varepsilon_{ij}=0}, \quad (19)$$

which is very lengthy for $n \geq 2$. If we described the skyrmion lattice phase within the triple-Q representation, we have $\mathbf{m} = \mathbf{m}_{tripleQ}(m_0, m_{q1}, q)$, where $q$ is found to be invariant, and the work conjugate of $m_{q1}$ is denoted by $b_{q1}$. In this case, eq. (19) reduces to

$$\left(\tilde{C}_{ijkl}\right)_{t,b,b_{q1},\varepsilon_{ij}=0} = \left[\frac{\partial(\tilde{\sigma}_{ij}, b, b_{q1})/\partial(\varepsilon_{kl}, m_0, m_{q1})}{\partial(\varepsilon_{kl}, b, b_{q1})/\partial(\varepsilon_{kl}, m_0, m_{q1})}\right]_{t,b,b_{q1},\varepsilon_{ij}=0}, \quad (20)$$

Substituting $\tilde{\sigma}_{ij} = \frac{\partial \bar{w}}{\partial \varepsilon_{ij}}$, $b = \frac{\partial \bar{w}}{\partial m_0}$, and $b_{q1} = \frac{\partial \bar{w}}{\partial m_{q1}}$ into eq. (20), after manipulation we have

$$\left(\tilde{C}_{ijkl}\right)_{t,b,b_{q1},\varepsilon_{ij}=0} = \frac{\partial^2 \bar{w}}{\partial \varepsilon_{ij} \partial \varepsilon_{kl}} + \frac{1}{\frac{\partial^2 \bar{w}}{\partial m_0^2}\frac{\partial^2 \bar{w}}{\partial m_{q1}^2} - \left(\frac{\partial^2 \bar{w}}{\partial m_0 \partial m_{q1}}\right)^2} \times \left[\frac{\partial^2 \bar{w}}{\partial \varepsilon_{ij} \partial m_0}\left(\frac{\partial^2 \bar{w}}{\partial m_0 \partial m_{q1}}\frac{\partial^2 \bar{w}}{\partial \varepsilon_{kl} \partial m_{q1}}\right.\right. \quad (21)$$
$$\left.\left. - \frac{\partial^2 \bar{w}}{\partial \varepsilon_{kl} \partial m_0}\frac{\partial^2 \bar{w}}{\partial m_{q1}^2}\right) + \frac{\partial^2 \bar{w}}{\partial \varepsilon_{ij} \partial m_{q1}}\left(\frac{\partial^2 \bar{w}}{\partial \varepsilon_{kl} \partial m_0}\frac{\partial^2 \bar{w}}{\partial m_0 \partial m_{q1}} - \frac{\partial^2 \bar{w}}{\partial m_0^2}\frac{\partial^2 \bar{w}}{\partial \varepsilon_{kl} \partial m_{q1}}\right)\right].$$

## IV. Results for bulk MnSi

MnSi is a prototype material for us to understand the interaction between magnetic skyrmions and mechanical loads. In this section, we apply the theory established above to bulk MnSi. Within a unified theoretical framework, we are able to quantitatively explains four different aspects of experimental results, including the temperature-magnetic field phase diagram for materials free from any mechanical loads, the variation of magnetostriction with magnetic field, the variation of elastic stiffness with magnetic field, and the temperature-magnetic field phase diagram for materials suffering uniaxial pressure. By doing so, a comprehensive set of thermodynamic parameters for MnSi is obtained and listed in Table 1, describing its magnetic, elastic, and magnetoelastic properties. The parameters are divided into two groups. The first group contains the parameters that can be directly obtained or have already been fitted from experiments, while the second group contains the parameters that are fitted in this work.

Table 1. Thermodynamic parameters for bulk $MnSi$

The first group of parameters:

$$C_{11} = 283.3 \text{ GPa}, C_{12} = 64.1 \text{ GPa}, C_{44} = 117.9 \text{ GPa [40]},$$

$$A = 1.27 \times 10^{-23} \text{ JA}^{-2}\text{m [5, 24, 41]}, D = 1.14 \times 10^{-14} \text{ JA}^{-2} \text{ [24, 42]}, M_s = 1.63 \times 10^5 \text{ A/m [24]}$$

$$\alpha = 6.44 \times 10^{-7} \text{ JA}^{-2}\text{m}^{-1}\text{K}^{-1}, \beta = 3.53 \times 10^{-16} \text{ JA}^{-4}\text{m}, T_0 = 26 \text{ K}, A_c = -0.05A, \text{ [36]}$$

The second group of parameters:

$$K = -2 \times 10^7 \text{ JA}^{-2}\text{m}^{-1}, L_1 = -0.70 \times 10^6 \text{ JA}^{-2}\text{m}^{-1}, L_2 = 0.60 \times 10^6 \text{ JA}^{-2}\text{m}^{-1}, L_3 = 1.65 \times 10^6 \text{ JA}^{-2}\text{m}^{-1},$$

$$L_{O1} = -0.57 \times 10^{-4} \text{ JA}^{-2}\text{m}^{-2}, L_{O2} = 1.15 \times 10^{-4} \text{ JA}^{-2}\text{m}^{-2}, L_{O3} = -0.57 \times 10^{-4} \text{ JA}^{-2}\text{m}^{-2},$$

$$L_{22} = -1.01 \times 10^8 \text{ JA}^{-2}\text{m}^{-1}, L_{23} = 2.03 \times 10^7 \text{ JA}^{-2}\text{m}^{-1}, B_c = 0.$$

We briefly introduce the method and the experimental data used to fit the second group of parameters. From magnetostriction experiment of bulk MnSi[28], $L_1, L_2, L_3$ and $L_{Oi}, (i = 1, 2, 3)$ can be fitted, where experimental results when the magnetic field is applied along three different directions (001), (110), and (111) are used. $K, L_{22}$ and $L_{23}$ are fitted from ultrasound measurements of the variation of the elastic coefficients with external magnetic field[27].

Using this set of thermodynamic parameters, extensive investigation is done concerning the magnetoelastic effects in MnSi, including the calculation of magnetostriction in the skyrmion phase and comparison to corresponding experiments, the calculation of the periodic elastic field in the skyrmion phase[29], the bumpy surface configuration of the skyrmion lattice[43], and the emergent elastic properties of the skyrmion lattice[38]. Here, we present three parts of calculation results for bulk MnSi.

**A. Temperature-magnetic field phase diagram concerning magnetoelastic coupling when the material is free from any mechanical loads**

Due to the magnetoelastic coupling, the elastic strains are related to the magnetization. Even when the system is free from any mechanical loads, we have nontrivial elastic strains from eq. (10). This nontrivial $\varepsilon_{ij}$ has an effect on the phase diagram as well as the solution of equilibrium magnetization through $\widetilde{w}_{el}$ and $\widetilde{w}_{me}$ in eq. (7). In phase diagram calculation of chiral magnets based on micromagnetic models[1, 35, 36], this back-action on the magnetization due to magnetoelastic coupling is usually neglected due to its smallness compared with other dominant terms in the free energy functional. To examine the applicability of such an assumption, we provide here a general analysis of the effect of magnetoelastic coupling on the magnetic phase diagram calculation for bulk chiral magnets when the system is free from any mechanical loads. We know that the coefficients of magnetoelastic coupling in different orders generally satisfy $\widetilde{K} \gg \widetilde{L}_1, \widetilde{L}_2, \widetilde{L}_3 \gg \widetilde{L}_{O1}q, \widetilde{L}_{O2}q, \widetilde{L}_{O3}q$, which gives $\varepsilon_{11}, \varepsilon_{22}, \varepsilon_{33} \propto K^*m^2$ and that the influence of elastic strains on the equilibrium magnetization is dominantly attributed to the term $\widetilde{K}m^2\varepsilon_{ii}$ in eq. (2). Hence, consideration of $\varepsilon_{11}, \varepsilon_{22}, \varepsilon_{33}$ in minimization of $\bar{w}$ renormalizes the coefficient of $m^4$ in the magnitude by $-\frac{\widetilde{K}^2}{\tilde{C}_{11}+2\tilde{C}_{12}}$. In summary, we can compare the value of $\left|\frac{\widetilde{K}^2}{\tilde{C}_{11}+2\tilde{C}_{12}}\right|$ and 1 to qulitatively evaluate if the coupling between the elastic strains and magnetization can be neglected in the phase diagram calculation. When $\left|\frac{\widetilde{K}^2}{\widetilde{C}_{11}+2\widetilde{C}_{12}}\right| \ll 1$, the equilibrium magnetization and magnetization induced elastic strains can be solved independently. For bulk $MnSi$, we have $\left|\frac{\widetilde{K}^2}{\widetilde{C}_{11}+2\widetilde{C}_{12}}\right| \sim 10^{-3}$, which suggests that magnetoelastic coupling should have a negligible effect on the shape of the magnetic phase diagram.

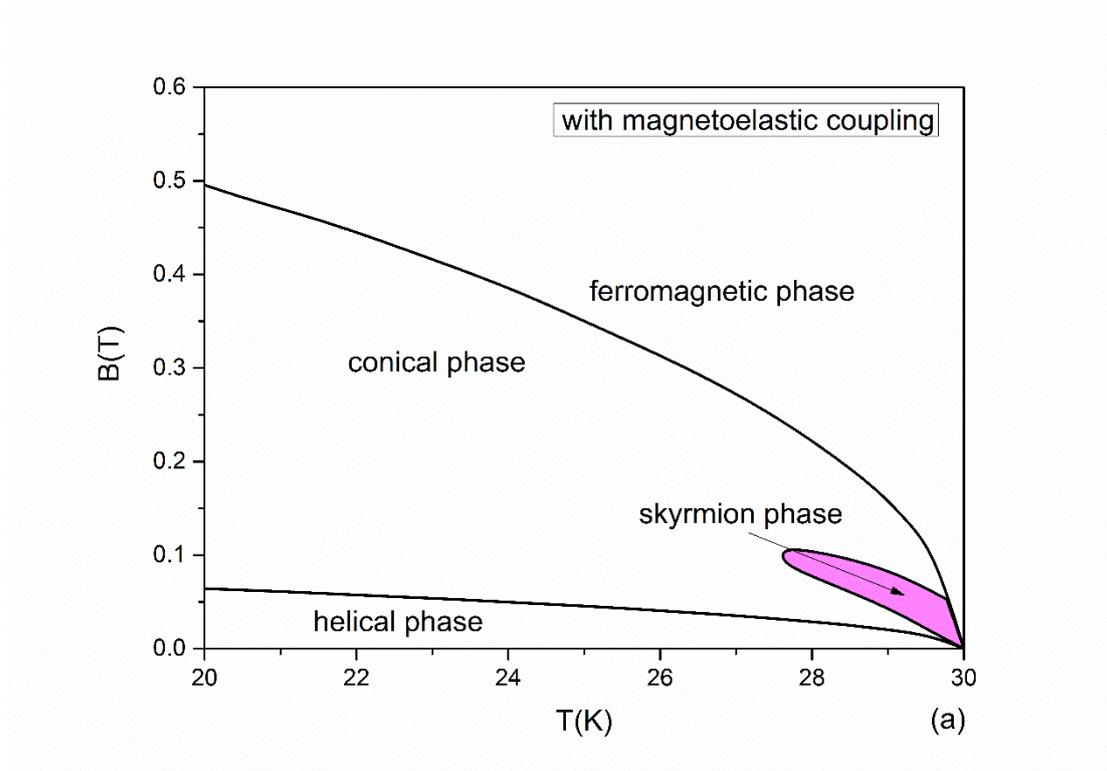

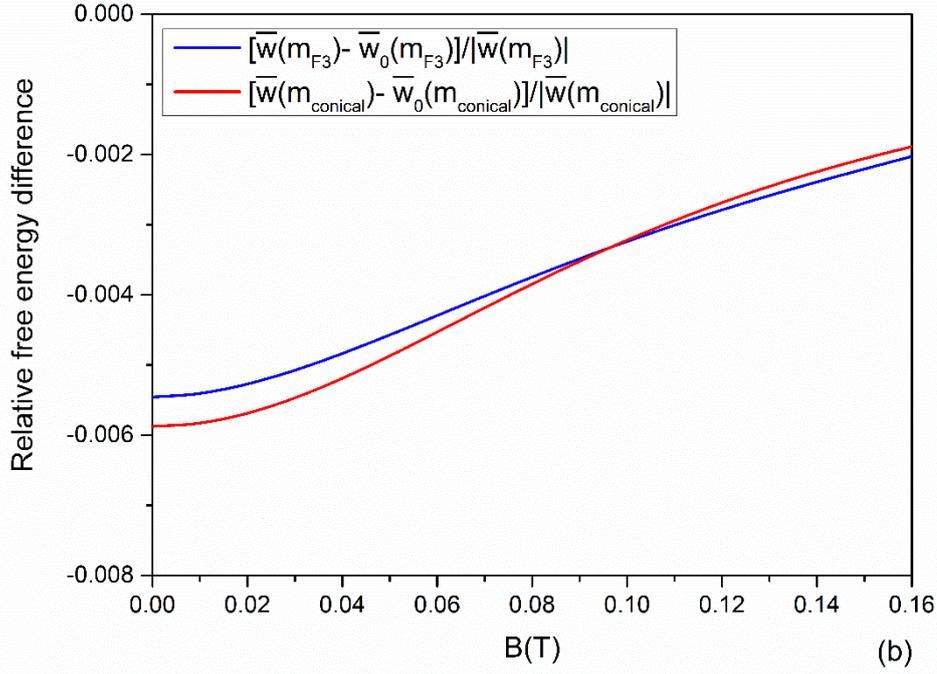

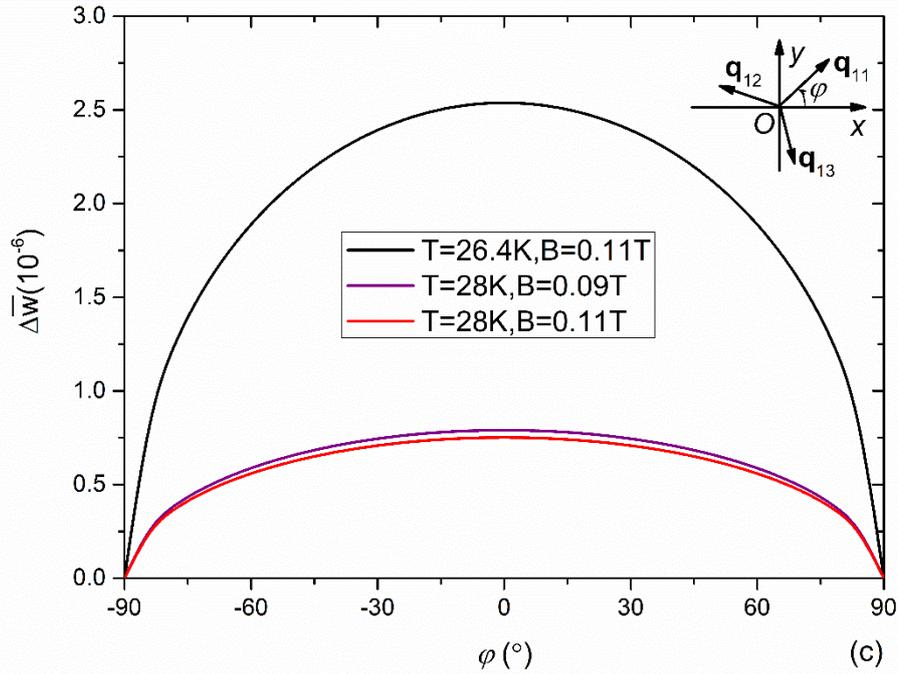

Figure 1. (a)Temperature-magnetic field phase diagram of MnSi when the system is free from any mechanical loads. (b) Relative free energy difference induced by magnetoelastic coupling as a function of the magnetic field calculated at temperature $T = 28$K. Here $\bar{w}(\mathbf{m}_{F3})$ and $\bar{w}(\mathbf{m}_{conical})$ denote the free energy calculated for the skyrmion phase and the conical phase incorporating magnetoelastic coupling at given magnetic field, while $\bar{w}_0(\mathbf{m}_{F3})$ and $\bar{w}_0(\mathbf{m}_{conical})$ denote the free energy calculated without considering the magnetoelastic coupling at given magnetic field. (c)

Variation of $\Delta \bar{w}$ with $\varphi$ at three conditions of $T$ and $B$.

For bulk MnSi free from any mechanical loads, we plot the temperature-magnetic field phase diagram in Figure 1(a) using the thermodynamic parameters fitted in Table 1. The phase diagram is indistinguishable from the one plotted in our previous work[36] where magnetoelastic coupling is neglected. To further understood why this is the case, we plot in Figure 1(b) the relative difference of free energy in the skyrmion phase $[\bar{w}(\mathbf{m}_{F3}) - \bar{w}_0(\mathbf{m}_{F3})]/|\bar{w}(\mathbf{m}_{F3})|$, and the relative difference of free energy in the conical phase $[\bar{w}(\mathbf{m}_{conical}) - \bar{w}_0(\mathbf{m}_{conical})]/|\bar{w}(\mathbf{m}_{conical})|$. Here $\bar{w}(\mathbf{m})$ and $\bar{w}_0(\mathbf{m})$ denote respectively the averaged free energy density calculated by integrating eq. (7) and the averaged free energy density calculated by neglecting $\widetilde{w}_{el} + \widetilde{w}_{me}$ in eq. (7). We see that at 28K, by including $\widetilde{w}_{el} + \widetilde{w}_{me}$ in the free energy density, the free energy of the system in both the skyrmion phase and the conical phase decreases by about 2% to about 6% at different magnetic field. Yet the free energy in the two phases decrease in approximately the same way, which explains why the phase diagram is merely affected. On the other hand, magnetoelastic coupling pins the direction of wave vectors ($\mathbf{q}_{11}$, $\mathbf{q}_{12}$, $\mathbf{q}_{13}$ and so on) in the skyrmion lattice phase, as shown in Figure 1(c). In the triple-Q representation of the skyrmion lattice, it is usually assumed that $\mathbf{q}_{11} = [1\ 0\ 0]^T$, $\mathbf{q}_{12} = [-1/2\ \sqrt{3}/2\ 0]^T$, and $\mathbf{q}_{13} = [-1/2\ -\sqrt{3}/2\ 0]^T$. This is because neglecting magnetoelastic coupling (setting $w_{el} = w_{me} = 0$ in eq. (7)), the free energy density functional is invariant under an arbitrary rotation of $\mathbf{q}_{11}$, $\mathbf{q}_{12}$ and $\mathbf{q}_{13}$ in the x-y plane. After incorporating magnetoelastic coupling in the free energy density, we find that such a symmetry is broken and the direction of the triple-Q wave vectors is pinned. In Figure 1(c), we plot in Figure 2(a) the variation of $\Delta \bar{w}[\mathbf{m}_{F3}(T,B,\varphi)] = \left[\bar{w}[\mathbf{m}_{F3}(T,B,\varphi)] - \bar{w}[\mathbf{m}_{F3}(T,B,\frac{\pi}{2})]\right]/\bar{w}[\mathbf{m}_{F3}(26.4, 0.11, \frac{\pi}{2})]$ with $\varphi$, where $\varphi$ describes the angle between $\mathbf{q}_{11}$ and the x-axis (depicted in the inset of Figure 2(a)). We find that at any temperature and magnetic field, the minimized free energy is found when $\varphi = \pm \frac{\pi}{2}$. As a result, magnetoelastic coupling pins the triple-Q wave vectors to $\mathbf{q}_{11} = [0\ 1\ 0]^T$, $\mathbf{q}_{12} = [-\sqrt{3}/2\ -1/2\ 0]^T$, and $\mathbf{q}_{13} = [\sqrt{3}/2\ -1/2\ 0]^T$, as is introduced in eq. (18). Meanwhile, it is found that the value of other independent variables such as $m_0, m_q,$ and $q$ in the triple-Q representation is merely affected by the magnetoelastic coupling.

## B. Variation of elastic constants with external magnetic field

Using the parameters given in Table 1 and the method introduced in section III part B, we plot the variation of elastic coefficients, denoted by $\Delta C_{\alpha\beta} = (C_{\alpha\beta})_{T,b_0,b_{i1},b_{i3},b_{i3}} - C_{\alpha\beta}$, $(\alpha, \beta = 1,2,...,6)$ in Figure 2(a-c). Here $(C_{\alpha\beta})_{T,b_0,b_{i1},b_{i3},b_{i3}}$ is derived from eq. (19), where $C_{\alpha\beta}$ denotes a compressed matrix notation of the fourth-order tensor $C_{ijkl}$[44]. The curves obtained from our theoretical calculation resemble corresponding experimental results in great detail[26, 27]. To be more specific, in Figure 2(a) $\Delta C_{11}$ and $\Delta C_{33}$ change in opposite direction from the same point in the distorted conical phase as the magnetic field increases[26, 27]; in the conical phase $\Delta C_{33} > \Delta C_{11}$ with a gap whose magnitude approaches $10^8$ GPa and the gap mildly decreases as the magnetic field increases[26]; when a phase transition from the conical phase to the skyrmion phase occurs, we observe an obvious lift of $\Delta C_{11}$ which makes $\Delta C_{11} > \Delta C_{33}$[26, 27]; when a phase transition from the conical phase to the ferromagnetic phase occurs, $\Delta C_{11}$ and $\Delta C_{33}$ both increases while $\Delta C_{11}$ increases more sharply[27]. In Figure 2(b), $\Delta C_{44}$ and $\Delta C_{66}$ change in opposite direction in both the distorted conical phase and the conical phase[27]; when a phase transition from the conical phase to the skyrmion phase occurs, $\Delta C_{44}$ slightly drops while $\Delta C_{66}$ increases shapely[27]. In Figure 2(c), variation of $\Delta C_{12}$ and $\Delta C_{13}$ with magnetic field is predicted, where corresponding experiments have never been performed before.

When plotting Figure 2, the magnetization in the skyrmion phase is described by the 3$^{rd}$ order Fourier representation. We find that if we use the triple-Q representation instead, the values of $\Delta C_{\alpha\beta}$ obtained slightly changes within a range of $\pm 0.2\%$. This result shows that the order of Fourier representation of the skyrmion lattice phase has a negligible effect on the calculation of elastic constants.

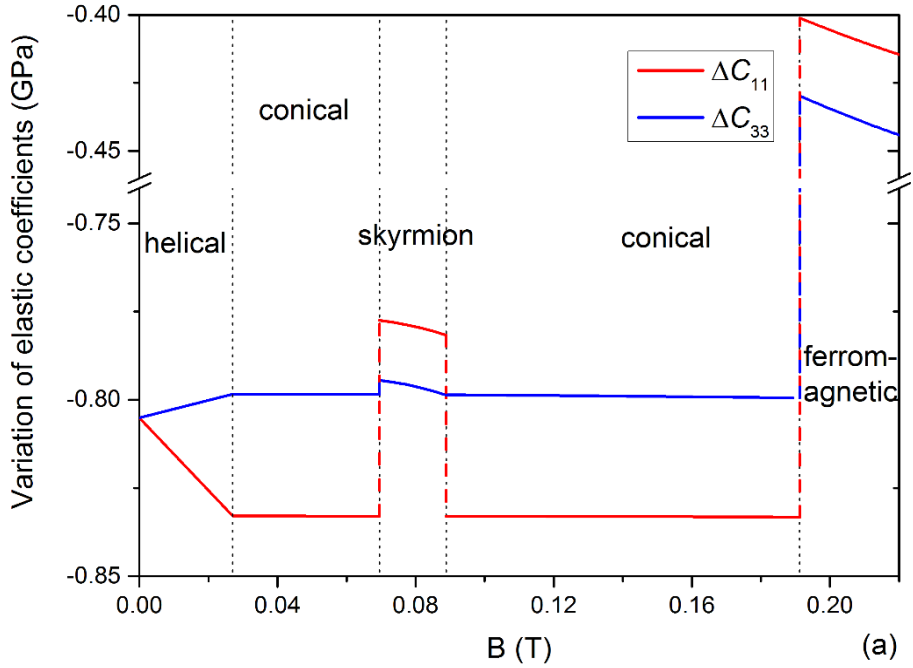

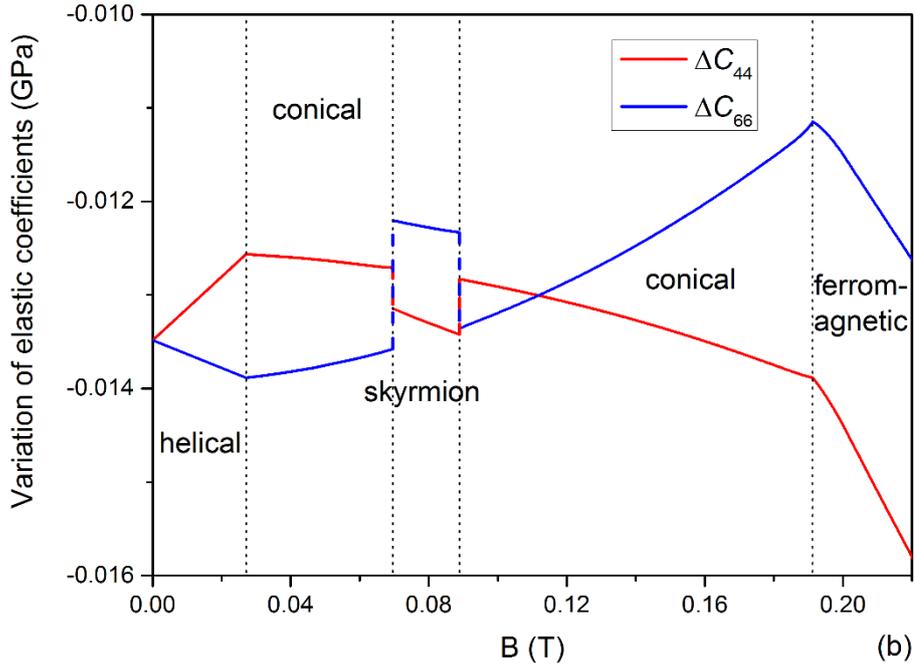

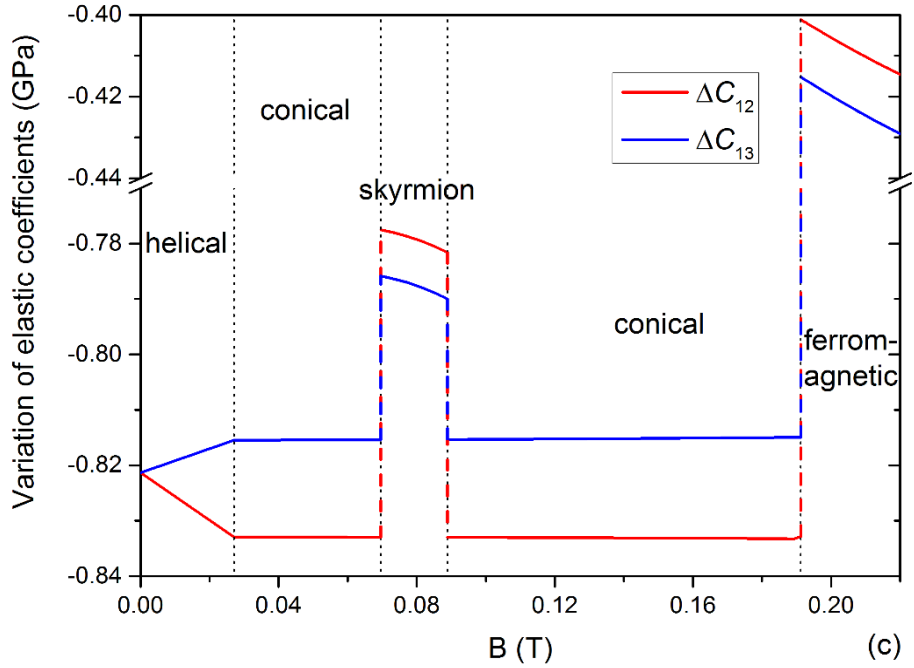

Figure 2. Variation of the elastic coefficients (a) $\Delta C_{11}, \Delta C_{33}$, (b) $\Delta C_{44}, \Delta C_{66}$, (c) $\Delta C_{12}, \Delta C_{13}$, of MnSi with external magnetic field at 28K. $\Delta C_{11}$ denotes the change of elastic coefficient $C_{11}$ due to magnetoelastic coupling.

## C. Temperature-magnetic field phase diagram for MnSi suffering uniaxial compression

We consider the variation of temperature-magnetic field phase diagram with application of uniaxial compression. The uniaxial compression is applied in two directions: $[0 \quad 0 \quad 1]^T$ and $[1 \quad 1 \quad 0]^T$. When uniaxial normal stress $\sigma$ is applied in $[0 \quad 0 \quad 1]^T$, the boundary condition reads $\sigma_{33} = \sigma$, $\sigma_{11} = \sigma_{22} = \sigma_{12} = \sigma_{13} = \sigma_{23} = 0$, and the elastic strains can be approximated by $\varepsilon_{11} = \varepsilon_{22} = -\frac{C_{12}}{C_{11}^2 + C_{11}C_{12} - 2C_{12}^2}\sigma$, $\varepsilon_{33} = \frac{C_{11} + C_{12}}{C_{11}^2 + C_{11}C_{12} - 2C_{12}^2}\sigma$, $\varepsilon_{12} = \varepsilon_{13} = \varepsilon_{23} = 0$. When uniaxial normal stress $\sigma$ is applied in $[1 \quad 1 \quad 0]^T$, the boundary condition reads $\sigma_{11} = \sigma_{22} = \sigma_{12} = \frac{\sigma}{2}, \sigma_{33} = \sigma_{13} = \sigma_{23} = 0$, and the elastic strains can be approximated by $\varepsilon_{11} = \varepsilon_{22} = \frac{C_{11}}{2(C_{11}^2 + C_{11}C_{12} - 2C_{12}^2)}\sigma$, $\varepsilon_{33} = -\frac{C_{12}}{C_{11}^2 + C_{11}C_{12} - 2C_{12}^2}\sigma$, $\varepsilon_{12} = \frac{\sigma}{2C_{44}}, \varepsilon_{13} = \varepsilon_{23} = 0$. By substituting the solution of elastic strains into eq. (7), we can initiate the temperature-magnetic field phase diagram calculation when uniaxial stress is applied. For two conditions of $\sigma$: $\sigma = -100MPa$ and $\sigma = -200MPa$, we plot in Figure 3 four temperature-magnetic field phase diagrams for MnSi. Our result agrees quantitatively with corresponding experiments[25]. The phase diagrams of MnSi under uniaxial pressure show the following characters: i) a left shift of critical temperature as the pressure increases. This effect is dominantly caused by the magnetoelastic coupling term $\widetilde{K}m^2\varepsilon_{ii}$ in eq. (2). According to the theory of elasticity, $\varepsilon_{ii} = \frac{1}{C_{11} + 2C_{12}}\sigma_{ii} = \frac{1}{C_{11} + 2C_{12}}\sigma$, where $\sigma_{ii}$ is a stress tensor invariant so that this result is valid for uniaxial compression applied in any direction. Hence $\widetilde{K}m^2\varepsilon_{ii}$ renormalizes the second order landau expansion term in eq. (7) as $(t + \frac{\widetilde{K}}{C_{11} + 2C_{12}}\sigma)m^2$, for which the rescaled curie temperature reduces from 1 to approximately $1 - \frac{\widetilde{K}}{C_{11} + 2C_{12}}\sigma$. For MnSi, uniaxial compression always decreases the curie temperature since $\widetilde{K} < 0$. ii) uniaxial compression in the direction of $[0 \quad 0 \quad 1]^T$ constricts the stable region of skyrmion phase in the phase diagram, while uniaxial compression in the direction of $[1 \quad 1 \quad 0]^T$ extends the stable region of skyrmion phase in the phase diagram. This effect is dominantly caused by the magnetoelastic coupling term $\widetilde{w}_{me02} = \widetilde{L}_1(m_1^2\varepsilon_{11} + m_2^2\varepsilon_{22} + m_3^2\varepsilon_{33})$ in eq. (2). To explain this, we compare the averaged free energy density in the skyrmion phase within triple-Q representation $\bar{w}(\mathbf{m}_{tripleQ}) = \bar{w}(m_0, m_{q1}, q)$ and the averaged free energy density in the conical phase $\bar{w}(\mathbf{m}_{conical}) = \bar{w}(m_3, m_q, q)$. At given condition of external field we have approximately $m_3 = m_0$ and $m_q = \sqrt{6}m_{q1}$, which yields $\Delta\bar{w}_{me02} = \bar{w}_{me02}(\mathbf{m}_{tripleQ}) - \bar{w}_{me02}(\mathbf{m}_{conical}) = -\frac{1}{4}\widetilde{L}_1(\varepsilon_{11} + \varepsilon_{22})m_q^2 + \frac{1}{2}\widetilde{L}_1\varepsilon_{33}m_q^2$ . Notice that for MnSi $\widetilde{L}_1 < 0$, out-of-plane uniaxial compression yields negative $\varepsilon_{33}$ and positive $\varepsilon_{11}, \varepsilon_{22}$, for which $\Delta\bar{w}_{me02} > 0$ so the skyrmion phase becomes less stable. On the other hand, in-plane uniaxial compression yields positive $\varepsilon_{33}$ and negative $\varepsilon_{11}, \varepsilon_{22}$, for which $\Delta\bar{w}_{me02} < 0$ so the skyrmion phase becomes more stable.

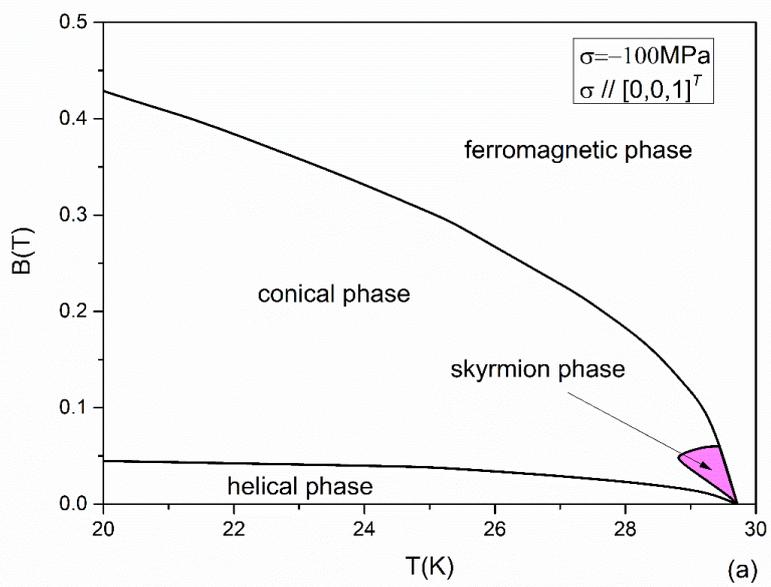

(a)

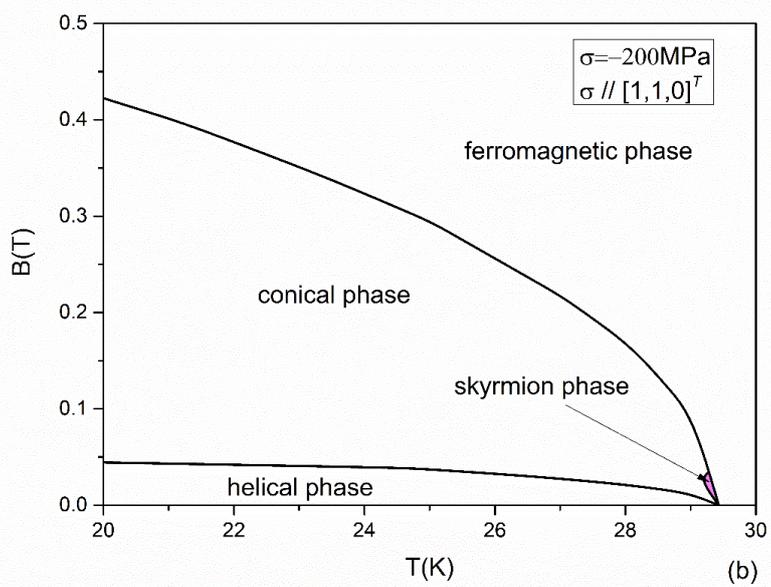

(b)

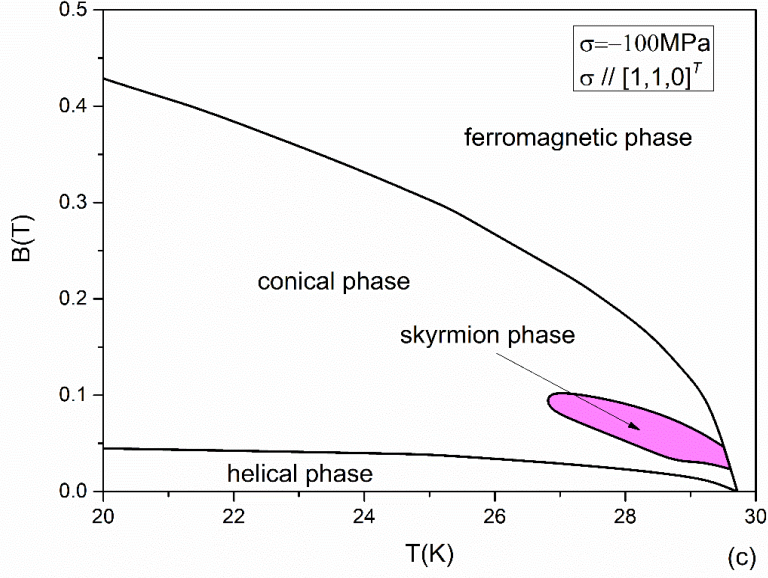

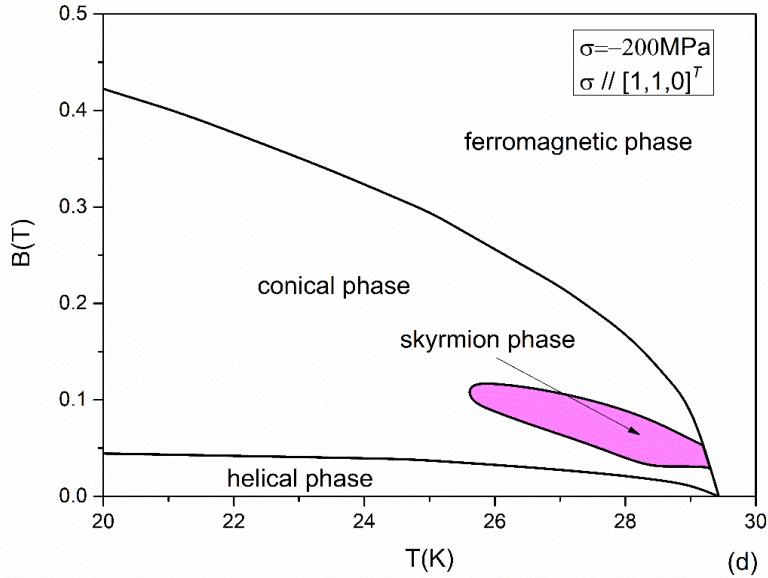

Figure 3. Temperature-magnetic field phase diagram of MnSi when the system is suffering uniaxial compression, where the compressive stress is applied in direction $[0\ \ 0\ \ 1]^T$ with (a) $\sigma = -100MPa$, (b) $\sigma = -200MPa$, and in direction $[1\ \ 1\ \ 0]^T$ with (c) $\sigma = -100MPa$, (d) $\sigma = -200MPa$.

**Appendix A. Simplification of the higher order magnetoelastic free energy density functional**

The lowest order coupling functional between $\varepsilon_{ij}, (i,j = 1,2,3)$, $m_l, (l = 1,2,3)$ and $m_{p,q}, (p,q = 1,2,3)$ that are invariant under all the operations of point group T can be written as:

$$\widetilde{w}_{me1T} = \sum_{i=1}^{13} \widetilde{L}_{Ti} \widetilde{f}_{Ti}, \tag{A1}$$

where

$$\begin{aligned}
\tilde{f}_{T1} &= \varepsilon_{11} m_{1,2} m_3 + \varepsilon_{22} m_{2,3} m_1 + \varepsilon_{33} m_{3,1} m_2, \\
\tilde{f}_{T2} &= \varepsilon_{11} m_{1,3} m_2 + \varepsilon_{22} m_{2,1} m_3 + \varepsilon_{33} m_{3,2} m_1, \\
\tilde{f}_{T3} &= \varepsilon_{11} m_{3,1} m_2 + \varepsilon_{22} m_{1,2} m_3 + \varepsilon_{33} m_{2,3} m_1, \\
\tilde{f}_{T4} &= \varepsilon_{11} m_{2,1} m_3 + \varepsilon_{22} m_{3,2} m_1 + \varepsilon_{33} m_{1,3} m_2, \\
\tilde{f}_{T5} &= \varepsilon_{11} m_1 m_{2,3} + \varepsilon_{22} m_2 m_{3,1} + \varepsilon_{33} m_3 m_{1,2}, \\
\tilde{f}_{T6} &= \varepsilon_{11} m_1 m_{3,2} + \varepsilon_{22} m_2 m_{1,3} + \varepsilon_{33} m_3 m_{2,1}, \\
\tilde{f}_{T7} &= \gamma_{23} m_{1,3} m_3 + \gamma_{13} m_{2,1} m_1 + \gamma_{12} m_{3,2} m_2, \\
\tilde{f}_{T8} &= \gamma_{23} m_{1,2} m_2 + \gamma_{13} m_{2,3} m_3 + \gamma_{12} m_{3,1} m_1, \\
\tilde{f}_{T9} &= \gamma_{23} m_{3,1} m_3 + \gamma_{13} m_{1,2} m_1 + \gamma_{12} m_{2,3} m_2, \\
\tilde{f}_{T10} &= \gamma_{23} m_{2,1} m_2 + \gamma_{13} m_{3,2} m_3 + \gamma_{12} m_{1,3} m_1, \\
\tilde{f}_{T11} &= \gamma_{23} m_1 m_{3,3} + \gamma_{13} m_2 m_{1,1} + \gamma_{12} m_3 m_{2,2}, \\
\tilde{f}_{T12} &= \gamma_{23} m_1 m_{2,2} + \gamma_{13} m_2 m_{3,3} + \gamma_{12} m_3 m_{1,1}, \\
\tilde{f}_{T13} &= \gamma_{23} m_1 m_{1,1} + \gamma_{13} m_2 m_{2,2} + \gamma_{12} m_3 m_{3,3}.
\end{aligned} \quad (A2)$$

Strictly speaking, point group T allows 13 independent thermodynamic parameters to describe $\widetilde{w}_{me1T}$. Yet in practice we do not have enough experimental data to fit all these parameters. Hence we change the symmetry condition from point group T to point group O, which yields $\widetilde{w}_{me1}$ defined in eqs. (3, 5). As for $\widetilde{w}_{me2}$, we neglect all terms that are relevant to $\varepsilon_{11}$, $\varepsilon_{22}$, $\varepsilon_{33}$, because $\widetilde{w}_{me2}$ describes higher order effects of those described by $\widetilde{w}_{me1}$. We find in calculation that these neglected terms relevant to $\varepsilon_{11}$, $\varepsilon_{22}$, $\varepsilon_{33}$ are related to the variation of elastic constants such as $C_{11}$, $C_{33}$. But their contribution is negligible compared with corresponding terms in $\widetilde{w}_{me1}$.

**Appendix B. Derivation of the rescaled free energy density functional incorporating magnetoelastic interactions**

The free energy density functional for cubic helimagnets incorporating magnetoelastic interactions can be written as

$$w(\mathbf{M}, \varepsilon_{ij}) = \sum_{i=1}^{3} A \left(\frac{\partial \mathbf{M}}{\partial x_i}\right)^2 + D\mathbf{M} \cdot (\nabla \times \mathbf{M}) - \mathbf{B} \cdot \mathbf{M} + \alpha(T - T_0)\mathbf{M}^2 + \beta \mathbf{M}^4 + w_{an} + w_{el} + w_{me}, \quad (B1)$$

where

$$w_{an} = \sum_{i=1}^{3} \left[ A_e \left(\frac{\partial M_i}{\partial x_i}\right)^2 + A_c M_i^4 \right], \quad (B2)$$

$$\begin{aligned}
w_{el} &= \frac{1}{2} C_{11}(\varepsilon_{11}^2 + \varepsilon_{22}^2 + \varepsilon_{33}^2) + C_{12}(\varepsilon_{11}\varepsilon_{22} + \varepsilon_{11}\varepsilon_{33} + \varepsilon_{22}\varepsilon_{33}) \\
&\quad + \frac{1}{2} C_{44}(\gamma_{12}^2 + \gamma_{13}^2 + \gamma_{23}^2),
\end{aligned} \quad (B3)$$

$$w_{me} = w_{me0} + w_{me1} + w_{me2}, \quad (B4)$$

and

$$\begin{aligned}
w_{me0} &= \frac{1}{M_s^2} [KM^2 \varepsilon_{ii} + L_1(M_1^2 \varepsilon_{11} + M_2^2 \varepsilon_{22} + M_3^2 \varepsilon_{33}) + L_2(M_3^2 \varepsilon_{11} + M_1^2 \varepsilon_{22} + M_2^2 \varepsilon_{33}) \\
&\quad + L_3 (M_1 M_2 \gamma_{12} + M_1 M_3 \gamma_{13} + M_2 M_3 \gamma_{23})],
\end{aligned} \quad (B5)$$

$$w_{me1} = \frac{1}{M_s^2} \sum_{i=1}^{6} L_{0i} f_{0i}, \qquad (B6)$$

$$w_{me2} = \frac{1}{M_s^2} \sum_{i=1}^{3} L_{2i} f_{2i}, \qquad (B7)$$

$$\begin{aligned}
f_{01} &= \varepsilon_{11}(M_{1,2}M_3 - M_{1,3}M_2) + \varepsilon_{22}(M_{2,3}M_1 - M_{2,1}M_3) + \varepsilon_{33}(M_{3,1}M_2 - M_{3,2}M_1), \\
f_{02} &= \varepsilon_{11}(M_{3,1}M_2 - M_{2,1}M_3) + \varepsilon_{22}(M_{1,2}M_3 - M_{3,2}M_1) + \varepsilon_{33}(M_{2,3}M_1 - M_{1,3}M_2), \\
f_{03} &= \varepsilon_{11}M_1(M_{2,3} - M_{3,2}) + \varepsilon_{22}M_2(M_{3,1} - M_{1,3}) + \varepsilon_{33}M_3(M_{1,2} - M_{2,1}), \\
f_{04} &= \gamma_{23}(M_{1,3}M_3 - M_{1,2}M_2) + \gamma_{13}(M_{2,1}M_1 - M_{2,3}M_3) + \gamma_{12}(M_{3,2}M_2 - M_{3,1}M_1), \\
f_{05} &= \gamma_{23}(M_{3,1}M_3 - M_{2,1}M_2) + \gamma_{13}(M_{1,2}M_1 - M_{3,2}M_3) + \gamma_{12}(M_{2,3}M_2 - M_{1,3}M_1), \\
f_{06} &= \gamma_{23}M_1(M_{3,3} - M_{2,2}) + \gamma_{13}M_2(M_{1,1} - M_{3,3}) + \gamma_{12}M_3(M_{2,2} - M_{1,1}),
\end{aligned} \qquad (B8)$$

$$\begin{aligned}
f_{21} &= \gamma_{23}\gamma_{12}M_1 M_3 + \gamma_{23}\gamma_{13}M_1 M_2 + \gamma_{12}\gamma_{13}M_2 M_3, \\
f_{22} &= \gamma_{23}^2 M_1^2 + \gamma_{12}^2 M_3^2 + \gamma_{13}^2 M_2^2, \\
f_{23} &= M^2(\gamma_{23}^2 + \gamma_{12}^2 + \gamma_{13}^2).
\end{aligned} \qquad (B9)$$

Here $\mathbf{M} = [M_1, M_2, M_3]^T$ denotes the magnetization vector, $M_s$ denotes the saturation magnetization, and $M^2 = M_1^2 + M_2^2 + M_3^2$. The first term in eq. (B1) describes the exchange energy density with stiffness $A$; the second term is the Zeeman energy density with the applied magnetic field $\mathbf{B}$; the third term is the Dzyaloshinskii-Moriya (DM) coupling with constant $b$ which determines the period and direction of the periodic magnetization; $\alpha(T - T_0)\mathbf{M}^2 + \beta \mathbf{M}^4$ are two Landau expansion terms. $w_{an}, w_{el}$ and $w_{me}$ denote respectively the anisotropy energy density with cubic magnetocrystalline anisotropic coefficient $A_c$ and exchange anisotripic coefficient $A_e$, the elastic energy density given in eq. (B3) and magnetoelastic free energy density given in eqs. (B4-B9).

Eq. (B1) provides an implicit model to study the effect of magnetoelastic interactions on the skyrmion lattice phase, because the effect cannot be understood by simply examining the magnetoelastic thermodynamic parameters, such as $K$, $L_1$, etc., but is also related to the magnetic thermodynamic parameters such as $A$ and $D$. In this case, it is more convenient to write the free energy density functional in a rescaled form given in eq. (7), where

$$\widetilde{w}(\mathbf{m}) = \frac{\beta}{G^2} w(\mathbf{M}), \quad (B10)$$

and

$$\mathbf{r} = \frac{\mathbf{x}}{L_D}, \mathbf{b} = \frac{\mathbf{B}}{B}, \mathbf{m} = \frac{\mathbf{M}}{M_0}, L_D = \frac{2A}{D}, G = \frac{D^2}{4A}, B = 2GM_0, M_0 = \sqrt{\frac{G}{\beta}}, t = \frac{\alpha(T - T_0)}{G}. \quad (B11)$$

The rescaled thermodynamic parameters (parameters with a wavy overline) are defined by

$$\widetilde{A}_e = \frac{A_e}{A}, \widetilde{A}_c = \frac{A_c}{\beta}, \widetilde{K} = \frac{K}{GM_s^2}, \widetilde{L}_1 = \frac{L_1}{GM_s^2}, \widetilde{L}_2 = \frac{L_2}{GM_s^2}, \widetilde{L}_3 = \frac{L_3}{GM_s^2}, \widetilde{L}_{2i} = \frac{L_{2i}}{GM_s^2}, (i = 1,2,3), \widetilde{L}_{0i}$$

$$= \frac{2L_{0i}}{DM_s^2}, (i = 1,2,\dots,6), \widetilde{C}_{11} = \frac{\beta}{K^2} C_{11}, \widetilde{C}_{12} = \frac{\beta}{K^2} C_{12}, \widetilde{C}_{44} = \frac{\beta}{K^2} C_{44}. \quad (B12)$$

The rescaled stress components are defined by $\widetilde{\sigma}_{ij} = \frac{\beta}{G^2} \sigma_{ij}$.

## V. Conclusion

In this paper, a thermodynamic model analyzing the coupled magnetoelastic fields in B20 helimagnets is developed based on group theoretical analysis. The model provides a unified theoretical framework to explain various aspects of skyrmion-related magnetoelastic experimental results, including but not limited to phase diagram calculation and equilibrium properties calculation under coupled temperature-magneto-elastic field. By applying the model to bulk MnSi, we quantitatively reproduce the temperature-magnetic field phase diagram when the material is free from any mechanical loads, the variation of all the elastic constants with magnetic field, and the variation of temperature-magnetic field phase diagram when the material is suffering uniaxial compression in two different directions. We also obtain the general condition at which the effect of magnetoelastic coupling on the equilibrium properties can be neglected, and find that magnetoelastic coupling pins the triple-Q wave vectors of the skyrmion lattice phase in the x-y plane. Through calculation, we fit a whole set of thermodynamic parameters for MnSi, which lays a reliable foundation for further analytical or numerical analysis of magnetoelastic coupling phenomena.

**Acknowledgement**: The work was supported by the NSFC (National Natural Science Foundation of China) through the fund 11302267, 11472313, 11572355.

**Author contributions**:
Y. Hu conceived the idea and finished the analytical deduction. Y. Hu and B. Wang discussed the results for revision. Y. Hu and B. Wang co-wrote the manuscript.

**Competing financial interests:** The authors declare no competing financial interests.